\documentclass[preprint,showpacs,preprintnumbers,amsmath,amssymb,aps]{revtex4-1}
\usepackage{graphicx}
\usepackage{dcolumn}
\usepackage{bm}

\begin{document}

\title{Study on the high spectral intensity at the Dirac energy of single-layer graphene on an SiC substrate}

\author{Jinwoong Hwang}
\author{Choongyu Hwang}\email{ckhwang@pusan.ac.kr}
\affiliation{Department of Physics, Pusan National University, Busan 46241, Korea}

\date{\today}

\begin{abstract}
We have investigated electron band structure of epitaxially grown graphene on an SiC(0001) substrate using angle-resolved photoemission spectroscopy. In single-layer graphene, abnormal high spectral intensity is observed at the Dirac energy whose origin has been questioned between in-gap states induced by the buffer layer and plasmaron bands induced by electron-plasmon interactions. With the formation of double-layer graphene, the Dirac energy does not show the high spectral intensity any longer different from the single-layer case. The inconsistency between the two systems suggests that the main ingredient of the high spectral intensity at the Dirac energy of single-layer graphene is the electronic states originating from the coupling of the graphene $\pi$ bands to the localized $\pi$ states of the buffer layer, consistent with the theoretical prediction on the presence of in-gap states.
\end{abstract}

%
%
%
%
%

\maketitle

\section{Introduction}

The charge neutral point $E_{\rm D}$, where conduction and valence bands touch at a single point, reveals an important aspect of the physics of graphene~\cite{Neto}. For example, when $E_{\rm D}$ is aligned to Fermi energy $E_{\rm F}$, graphene shows strong electronic correlations that are not explained by the theory that describes typical metallic systems~\cite{HwangSR}. Upon introducing dopants (or charges), many-body effects are developed to show well-defined quasiparticle states near $E_{\rm D}$ induced by electron-plasmon interactions, so-called plasmaron bands, right next to the well-known graphene $\pi$ bands~\cite{MacDonald1, MacDonald2, Eli1, Eli2}. Such electronic correlations are strongly affected by the presence of a substrate. Especially, when dielectric screening from the substrate becomes stronger compared to that for free-standing graphene, the plasmaron bands approach towards the graphene $\pi$ bands, so that they are not resolved, but leave their signature as high spectral intensity at $E_{\rm D}$~\cite{Eli1, Eli2}. 

Alternatively the electron band structure itself can result in the spectral feature at $E_{\rm D}$. On the surface of an  SiC(0001) substrate, a carbidic layer, so-called buffer layer, is formed with the same geometric structure as graphene in the absence of the characteristic conical dispersion due to the formation of covalent bonds with the substate~\cite{Varchon}. Upon the formation of single-layer graphene on top of it, the presence of the buffer layer breaks the sublattice symmetry of the single-layer graphene resulting in an energy gap at $E_{\rm D}$~\cite{Kim,Lee,Zhou07,Rusponi}. Meanwhile localized $\pi$ states of the buffer layer couple to the graphene $\pi$ bands which contributes to finite spectral intensity at $E_{\rm D}$, i.\,e.\,, in-gap states, despite the plasmaron bands are not taken into account~\cite{Kim}. In fact, recent {\em GW} calculations give no well-defined plasmaron bands in the epitaxial doped single-layer graphene on an SiC substrate~\cite{Louie}, in contrast to the previous theoretical \cite{MacDonald1, MacDonald2} and experimental results~\cite{Eli1, Eli2}.

Here we investigate this controversial issue by comparing the electron band structure of single- and double-layer graphene samples using angle-resolved photoemission spectroscopy (ARPES). Single-layer graphene shows high spectral intensity at $E_{\rm D}$ consistent with previous results~\cite{Eli2,Zhou07}. On the other hand, in double-layer graphene, we observe that the electron band structure shows a dip in spectral intensity at $E_{\rm D}$. The difference in the spectral feature is well described in terms of the in-gap states, whereas the plasmaron bands does not explain the weak spectral intensity observed at $E_{\rm D}$ of double-layer graphene on an SiC(0001) substrate and even single-layer graphene on metallic substrates. Our results suggest that the in-gap states induced by the buffer layer~\cite{Kim} are the main ingredient of the controversial high spectral intensity at $E_{\rm D}$ of single-layer graphene on the SiC substrate. 

\section{Experimental methods}

Graphene samples were prepared by the epitaxial growth method on an SiC(0001) substrate~\cite{Rolling} and the chemical vapour deposition method on a Cu film~\cite{Hong}. High-resolution ARPES experiments have been performed on the graphene samples using a synchrotron source with an energy of 50~eV at beamline 12.0.1 of the Advanced Light Source at Lawrence Berkeley National Laboratory. The energy and angular resolutions are 32~meV and $\leq$0.3$^{\circ}$, respectively. All the measurements have been performed at 15~K in ultra-high vacuum with a base pressure of 3.5$\times$10$^{-11}$~Torr.

\section{Results and Discussions}

\begin{figure}[t]
\centering
\includegraphics[width=0.5\columnwidth]{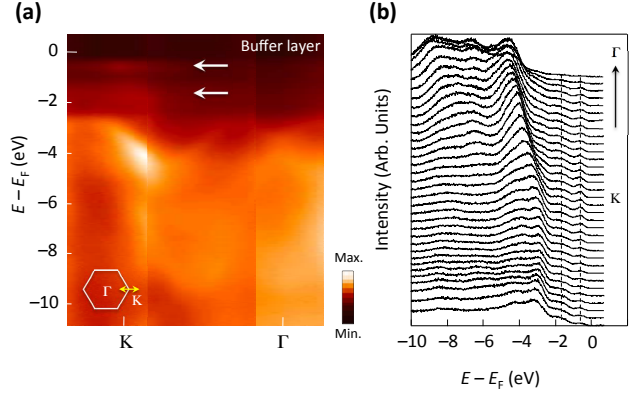}
\caption{\label{fig:fig1} (a) ARPES intensity map of the buffer layer sample along the $\Gamma-{\rm K}$ direction of graphene unit cell denoted by the yellow arrow in the inset. The buffer layer exhibits two non-dispersive states at 0.5~eV and 1.7~eV below Fermi energy, as denoted by white arrows. (b) EDCs, taken at different momentum positions for the ARPES map shown in panel (a), also show the non-dispersive feature as denoted by dashed lines.}
\end{figure}

Figure~\ref{fig:fig1}(a) shows an ARPES intensity map of a buffer layer sample taken along the $\Gamma-{\rm K}$ direction of the graphene unit cell denoted in the inset, which is consistent with the previous results~\cite{Emtsev}. More specifically, near $E_{\rm F}$ two non-dispersive states are observed at 1.7~eV and 0.5~eV below $E_{\rm F}$ as denoted by white arrows. Such a non-dispersive feature is also clear in energy distribution curves (EDCs) taken at different momentum values as denoted by dashed lines in Fig.~\ref{fig:fig1}(b). The former originates from the Si dangling bond states which belong to the Si-rich surface of the SiC(0001) substrate, so-called $\sqrt{3} \times \sqrt{3}$ phase, formed before the buffer layer is grown on the substrate~\cite{Emtsev}. With the formation of the buffer layer, several flat bands are predicted below $E_{\rm F}$ originating from the $\pi$-orbitals of carbon atoms~\cite{Kim}, which are observed as the non-dispersive state 0.5~eV below $E_{\rm F}$. The observation of the two non-dispersive states suggests that the sample is not fully covered by the buffer layer, but consists of both buffer layer and the $\sqrt{3} \times \sqrt{3}$ phase. 

\begin{figure*}
\centering
\includegraphics[width=1\columnwidth]{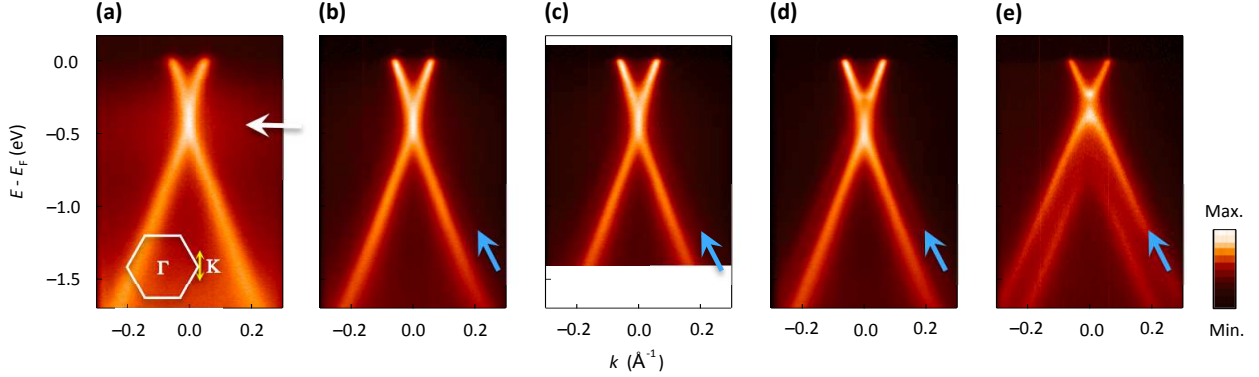}
\caption{\label{fig:fig2} (a)-(e) ARPES intensity maps of graphene samples near $E_{\rm F}$ taken perpendicular to the $\Gamma-{\rm K}$ direction denoted by the yellow arrow in the inset of panel (a). The white arrow denotes the non-dispersive state observed from the buffer layer sample shown in Fig.~\ref{fig:fig1}(a) and the blue arrows denote the evolution of the electron band structure corresponding to double-layer graphene.}
\end{figure*}

To investigate the latter non-dispersive state, we have measured ARPES intensity maps of several different graphene samples taken perpendicular to the $\Gamma-{\rm K}$ direction of graphene unit cell as denoted by the yellow arrow in the inset. Figure~\ref{fig:fig2}(a) shows the $\pi$ bands of single-layer graphene that are intrinsically doped by electrons due to the presence of the SiC substrate~\cite{Seyller}, so that $E_{\rm D}$ lies $\sim$0.43~eV below $E_{\rm F}$. The measured ARPES intensity map exhibits two additional features in addition to the characteristic conical dispersion: (i) finite spectral intensity away from the graphene $\pi$ bands as denoted by the white arrow; (ii) strong spectral intensity at $E_{\rm D}$ in the graphene $\pi$ bands. The former is the non-dispersive state observed from the buffer layer in Fig.~\ref{fig:fig1}. Coexistence of the non-dispersive state and the graphene $\pi$ bands suggests that the sample consists of single-layer graphene as well as the buffer layer, consistent with previous microscopy results~\cite{Siegel}. The origin of the latter has been argued between the in-gap states~\cite{Kim} and the plasmaron bands~\cite{Eli1}. As we go from the sample shown in Fig.~\ref{fig:fig2}(a) to Fig.~\ref{fig:fig2}(d), spectral weight of another band structure develops as denoted by the blue arrows, until it becomes very clear in Fig.~\ref{fig:fig2}(e) to show the characteristic electron band structure of double-layer graphene.

\begin{figure}[t]
\centering
\includegraphics[width=0.5\columnwidth]{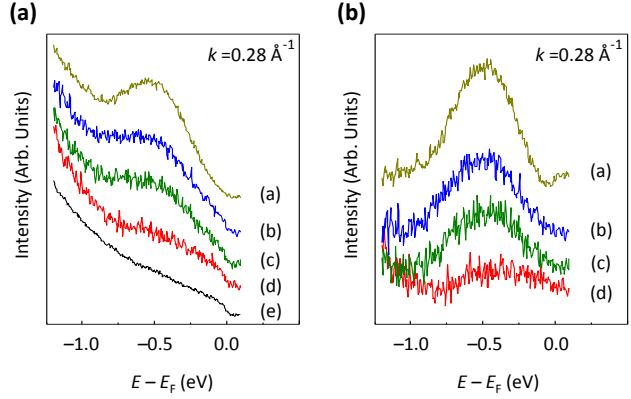}
\caption{\label{fig:fig3} (a) EDCs taken at $k=0.28~{\rm \AA}^{-1}$ from Figs.~\ref{fig:fig2}(a)-\ref{fig:fig2}(e). 
(b) The EDCs after subtracting background taken from the spectrum of the double-layer graphene, (e), shown in panel (a).}
\end{figure}

It is interesting to note that the finite spectral intensity denoted by the white arrow in Fig.~\ref{fig:fig2}(a) becomes weaker as we go towards the double-layer graphene sample shown in Fig.~\ref{fig:fig2}(e), which is obvious when we take EDCs at $k=0.28~{\rm \AA}^{-1}$ for each sample. Figure~\ref{fig:fig3}(a) shows normalized EDCs, when (a)-(e) denote that each spectrum is taken from Fig.~\ref{fig:fig2}(a)-(e), respectively. Figure~\ref{fig:fig3}(b) shows the normalized EDCs after subtracting spectral intensity of double-layer graphene denoted by (e) in Fig.~\ref{fig:fig2}(a). The intensity of the well-defined peak shape at $E-E_{\rm F}=-0.5~{\rm eV}$ decreases as we go from the sample shown in Fig.~\ref{fig:fig2}(a) to Fig.~\ref{fig:fig2}(d), which becomes featureless for the double-layer graphene sample (e).  It is known that, as the graphene layers develop, the buffer layer is increasingly covered by the graphene overlayers~\cite{Siegel}. The previous first principles calculations show that the localized $\pi$ states of the buffer layer couple to the graphene $\pi$ bands giving rise to finite spectral intensity at $E_{\rm D}$ of single-layer graphene on an SiC(0001) substrate. Within this theory, upon the evolution of graphene overlayers, the buffer layer states are increasingly coupled to the graphene $\pi$ bands, resulting in gradual decrease of spectral intensity of the buffer layer states. This is consistent with what we have observed in Figs.~\ref{fig:fig2} and~\ref{fig:fig3}.

Alternatively, due to the finite mean free path of electrons within a solid, increasing thickness of graphene layers can reduce photoelectron intensity from the buffer layer states. Indeed, as we go from Fig.~\ref{fig:fig2}(a) to Fig.~\ref{fig:fig2}(e), graphene layers on top of the buffer layer become thicker from single-atomic layer to 0.35~nm corresponding to double-layer graphene~\cite{Varchon}. This is comparable to a mean free path of $\sim$0.6~nm corresponding to a photon energy of 50~eV, that we have used in our experiments. However, since single-layer of MoSe$_2$ whose thickness (0.33~nm~\cite{Ugeda}) is comparable to that of double-layer graphene clearly shows photoelectron signal from its substrate~\cite{Zhang}, the mean free path issue for the samples with a thickness of $\leq$0.35~nm (Fig.~\ref{fig:fig2}) may be less likely the origin of the completely disappeared intensity of the buffer layer states observed from the double-layer graphene sample as shown in Figs.~\ref{fig:fig2}(e) and~\ref{fig:fig3}(a).

\begin{figure}[t]
\centering
\includegraphics[width=0.5\columnwidth]{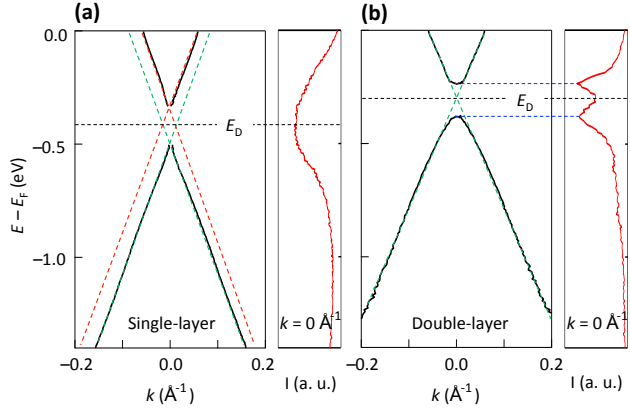}
\caption{\label{fig:fig4} (a) Fitted energy-momentum dispersion of the ARPES intensity map shown in Fig.~\ref{fig:fig2}(a). The green and red dashed lines are arbitrary straight lines for the guide to the eyes. The right panel is an EDC taken at $k=0~{\rm \AA}^{-1}$. Dirac energy is denoted by $E_{\rm D}$. (b) Fitted energy-momentum dispersion of the ARPES intensity map shown in Fig.~\ref{fig:fig2}(e). The green dashed lines are arbitrary straight lines for the guide to the eyes and the blue dashed lines denote the conduction band minimum and the valence band maximum. The right panel is an EDC taken at $k=0~{\rm \AA}^{-1}$.}
\end{figure}

The comparison between the energy spectra of single- and double-layer graphene provides an intriguing insight on the origin of the high spectral intensity at $E_{\rm D}$ for single-layer graphene. For example, Fig.~\ref{fig:fig4} shows fitted energy-momentum dispersions near $E_{\rm F}$ and EDCs for single- and double-layer graphene. In Fig.~\ref{fig:fig4}(a), the black line is the fitted graphene $\pi$ bands using a Lorentzian peak function, while the green  and red dashed lines are arbitrary straight lines along the fitted bands. At $E_{\rm D}$, where the extended lines meet, the EDC taken at $k=0~{\rm \AA}^{-1}$ shows maximum intensity. One of its plausible origins is the formation of plasmaron bands induced by electron-plasmon interactions~\cite{MacDonald1}. When graphene is placed on an SiC(0001) substrate, the dielectric screening can suppress these electronic correlations, so that the plasmaron bands approach towards the graphene $\pi$ bands~\cite{Eli1,Eli2}. As a result, the plasmaron bands might not be resolved with the typical experimental resolution, whereas their formation is evidenced as high spectral intensity at $E_{\rm D}$, similar to what we have observed in Fig.~\ref{fig:fig4}(a)~\cite{Eli1}.

\begin{figure}[t]
\centering
\includegraphics[width=0.5\columnwidth]{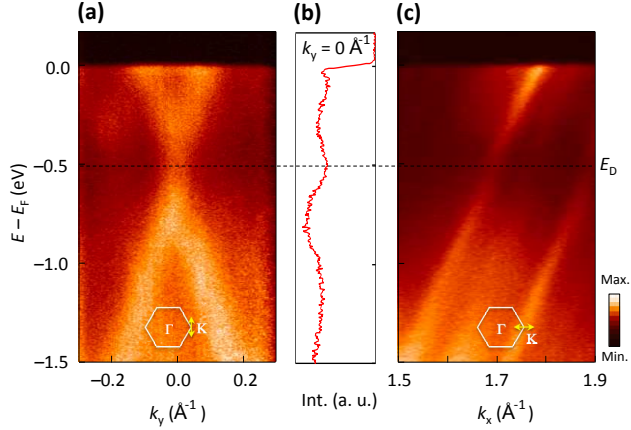}
\caption{\label{fig:fig5} (a) An ARPES intensity map of single-layer graphene on a Cu film near $E_{\rm F}$ taken perpendicular to the $\Gamma-{\rm K}$ direction denoted by the yellow arrow in the inset. $E_{\rm D}$ is denoted by the black dashed line. (b) An EDC taken at $k=0~{\rm \AA}^{-1}$. (c) An ARPES intensity map of single-layer graphene on a Cu film near $E_{\rm F}$ taken parallel to the $\Gamma-{\rm K}$ direction denoted by the yellow arrow in the inset.}
\end{figure}

Interestingly, however, single-layer graphene on metallic substrates~\cite{Varykhalov,Vita}, i.\,e.\,, the case where dielectric screening is further enhanced, does not show the characteristic feature of the screened plasmaron bands, but exhibits a dip at $E_{\rm D}$ in the EDC. Figure~\ref{fig:fig5} shows ARPES intensity maps of single-layer graphene on a Cu film. Along both perpendicular (Fig.~\ref{fig:fig5}(a)) and parallel (Fig.~\ref{fig:fig5}(c)) to the $\Gamma-{\rm K}$ direction of the graphene unit cell denoted in the inset, the graphene $\pi$ bands show weak spectral intensity (or a dip in the EDC) at $E_{\rm D}$ in contrast to the high spectral intensity observed at $E_{\rm D}$ of single-layer graphene on an SiC substrate.

Instead, the spectral feature observed from graphene on metallic substrates~\cite{Varykhalov,Vita} resembles that of double-layer graphene. At $E_{\rm D}$, determined by the crossing point of the green dashed lines drawn along the fitted bands in Fig.~\ref{fig:fig4}(b), the EDC shows a dip in between two peaks corresponding to the conduction band minimum and the valence band maximum (as denoted by blue dashed lines). This indicates that our double-layer graphene sample does not show the characteristic feature of the plasmaron bands reported in the case of single-layer graphene~\cite{MacDonald1,MacDonald2,Eli1,Eli2}. However, not only single-layer graphene, but also double-layer graphene is predicted to exhibit the plasmaron bands in the presence of charge imbalance between the adjacent layers~\cite{Phan}, which is intrinsically induced in epitaxial double-layer graphene on an SiC(0001) substrate~\cite{Ohta} as shown in Fig.~\ref{fig:fig2}(e). In addition, for single-layer graphene, the separation between the extended lines and the conduction (or valence) band is 95$\pm$14~meV, which is bigger than our experimental resolution, 32~meV, as much as by a factor of 3. This means that the additional feature, if any, e.\,g.\,, the plasmaron bands~\cite{MacDonald1,MacDonald2}, should be resolved within our experimental resolution. In reality however, we do not observe two separated bands corresponding to the graphene $\pi$ bands and plasmaron bands. These two observations cast doubt on the description of the high spectral intensity at $E_{\rm D}$ of single-layer graphene on an SiC(0001) substrate in terms of the plasmaron bands~\cite{Eli1}. In fact, a recent theoretical study suggests the absence of the well-defined plasmaron bands in the epitaxial doped single-layer graphene on the SiC substrate~\cite{Louie}.

The high spectral intensity at $E_{\rm D}$ can be alternatively explained by the in-gap states accompanied by the energy gap opening. The presence of the buffer layer under single-layer graphene can break sublattice symmetry resulting in the energy gap opening at $E_{\rm D}$~\cite{Zhou07}. At the same time, the localized $\pi$ states of the buffer layer are predicted to couple with the graphene $\pi$ bands resulting in finite spectral intensity in the gap region of single-layer graphene~\cite{Kim}. Within this picture, the spectral intensity of the localized $\pi$ states is expected to gradually decrease with the evolution of graphene overlayers, consistent with our observations shown in Figs.~2 and 3. In addition, when $E_{\rm D}$ for single- and double-layer graphene is different, the spectral intensity of the in-gap states is not expected to contribute to the spectral intensity at $E_{\rm D}$ of double-layer graphene. In fact, $E_{\rm D}$ is 0.43~eV and 0.30~eV below $E_{\rm F}$ for single- and double-layer graphene, respectively, as shown in Figs.~\ref{fig:fig4}(a) and~\ref{fig:fig4}(b), and the in-gap states (or the high spectral intensity) lies near 0.43~eV (Fig.~\ref{fig:fig4}(a)) below $E_{\rm F}$ which is close to the valence band maximum of double-layer graphene denoted by the lower blue dashed line in Fig.~\ref{fig:fig4}(b). These are exactly what we have manifested in Figs.~\ref{fig:fig2}-\ref{fig:fig4}.

\section{Summary}
We have investigated electron band structure of buffer layer, single- and double-layer graphene samples epitaxially grown on an SiC(0001) substrate. The spectral intensity of the non-dispersive states originating from the localized $\pi$ states of the buffer layer decreases with the evolution of double-layer graphene. At the same time, we found that the high spectral intensity at the Dirac energy is observed only for single-layer graphene, whereas double-layer graphene exhibits a dip in spectral intensity in between the conduction band minimum and valence band maximum. Our results suggest that the high spectral intensity observed at $E_{\rm D}$ of single-layer graphene is well described by the in-gap states picture, in which the buffer layer states are coupled to the graphene $\pi$ bands giving rise to a finite contribution in the spectral intensity at $E_{\rm D}$~\cite{Kim}, while the plasmaron bands picture does not explain different spectral intensity observed at $E_{\rm D}$ of single- and double-layer graphene.

\acknowledgements
This work was supported by Basic Science Research Program through the National Research Foundation of Korea (NRF) funded by the Ministry of Science, ICT \& Future Planning (No.~2015R1C1A1A01053065).

\end{document}